\documentclass{llncs}
\usepackage{amsmath,amsfonts,amssymb,url}

\begin{document}

\def\Real{\mathbb{R}} 

\def\Rat{\mathbb{Q}} 

\def\Zahl{\mathbb{Z}} 

\def\Par{\mathcal{P}} 

\def\Acc{\Gamma} 

\def\Forb{\mathcal{A}}  

\def\Auth{\mathcal{B}} 

\def\Af{F} 

\def\gap{g} 

\def\Fams{\mathcal{C}} 

\def\Poly{\mathcal{S}} 

\def\Ipoly{\mathcal{Z}} 

\def\Bas{\mathcal{B}} 

\def\Mat{\mathcal{M}} 

\def\Ind{\mathcal{I}} 

\def\cA{\mathcal{A}}

\def\kos{\mathbb{K}}

\def\maxp{\sigma} 
\def\bbound{\kappa} 
\def\lbound{\lambda} 

\def\vectx{\mathbf{x}}
\def\vecty{\mathbf{y}}
\def\vectu{\mathbf{u}}
\def\vectv{\mathbf{v}}
\def\vecte{\mathbf{e}}
\def\vects{\mathbf{s}}

\def\alice{\mathcal{C}}
\def\bob{\mathcal{S}}

\def\Ics{\mathbf{X}}
\def\Qu{\mathbf{Q}}
\def\Pe{\mathbf{P}}

\newcommand{\be}{{\mathbf{e}}}

\title{Flexible and Robust Privacy-Preserving Implicit Authentication}

\author{Josep Domingo-Ferrer\inst{1} \and Qianhong Wu\inst{2} 
\and Alberto Blanco-Justicia\inst{1}}

\institute{Universitat Rovira i Virgili\\
Dept. of Computer Engineering and Mathematics\\
UNESCO Chair in Data Privacy\\
Av. Pa\"{\i}sos Catalans 26, E-43007 Tarragona, Catalonia\\
\email{\{josep.domingo,alberto.blanco\}@urv.cat}
\and
School of Electronics and Information Engineering\\
Beihang University\\
XueYuan Road No. 37, Haidian District, Beijing, China\\
\email{qianhong.wu@buaa.edu.cn}}

\maketitle

\begin{abstract}
Implicit authentication consists of a server 
authenticating a user based on the user's usage profile,
instead of/in addition to relying on something the user explicitly
knows (passwords, private keys, etc.).
While implicit authentication makes identity theft 
by third parties more difficult, it requires the server
to learn and store the user's usage profile. 
Recently, the first privacy-preserving implicit authentication
system was presented, in which the server
does not learn the user's profile. It uses an {\em ad hoc} two-party
computation protocol to compare the user's fresh
sampled features
against an encrypted stored user's profile. The protocol requires storing
the usage profile and comparing against it using two different
cryptosystems, one of them order-preserving; 
furthermore, features must be numerical.
We present here a simpler protocol based on set intersection
that has the advantages of: i) requiring only one cryptosystem; ii)
not leaking the relative order of fresh feature samples; 
iii) being able to deal with any type of features (numerical or non-numerical).\\

\end{abstract}

\section{Introduction}

The recent report~\cite{ftc2014} by the U.S. 
Federal Trade Commission calls for transparency and 
accountability of data brokers. On the one hand, the report describes
the pervasive data collection carried
out by data brokers as clearly privacy-invasive.
On the other hand, it presents risk mitigation services offered
by data brokers as the good side of data collection, to the extent
that such services protect consumers against identity theft.
Indeed, storing information on how a consumer usually interacts
with a service (time of the day, usual places, usual sequence
of keystrokes, etc.) allows using this information to {\em implicitly
authenticate} a user: implicit authentication~\cite{Jako09} 
(a.k.a. transparent authentication~\cite{Clar09} 
or active authentication~\cite{Aksa09}) is the process
of comparing the user's current usage profile with the stored
profile. If both profiles disagree, maybe someone is impersonating
the user, {\em e.g.} after some identity theft (password theft, etc.).

The above risk mitigation argument is part 
a long-standing simplistic tendency in digital services (and elsewhere)  
to justify privacy invasion
in the name of legitimate interests, as if the latter
were incompatible with privacy (another old example
is intellectual property protection, which
was portrayed as being incompatible with 
the anonymity of digital content consumers
until anonymous fingerprinting 
was proposed~\cite{Pfit97,Domi98,Megi14}). 
In fact, implicit authentication turns out to be a weak
excuse to justify the storage and/or access by 
servers to the usage profiles of users.
In~\cite{Safa14} it was shown how to make implicit authentication
compatible with the privacy of users. 
The idea is that the server
only needs an {\em encrypted} version of the user's usage profile.

\subsection{Contribution and plan of this paper}

The protocol
in~\cite{Safa14} needs the server to store the 
users' accumulated usage profiles encrypted under {\em two} different
cryptosystems, one that is homomorphic and one that is order-preserving.
 We present here a protocol for privacy-preserving implicit
authentication based on set intersection,  which has the advantage
that the server only needs to store the users' accumulated usage profiles 
encrypted under {\em one} (homomorphic) cryptosystem.
This allows saving storage at the carrier and also computation
during the protocol.
Also, unlike~\cite{Safa14}, our protocol does not leak
the relative order of fresh feature samples collected by 
the user's device for comparison with the encrypted profile.
Finally, our protocol can deal with any type of features
(numerical or non-numerical), while the protocol~\cite{Safa14}
is restricted to numerical features.

The rest of this paper is organized as follows.
Section~\ref{back} gives background on implicit authentication
and on the privacy-preserving implicit authentication protocol 
of~\cite{Safa14}.
Section~\ref{dist} discusses how to compute the
dissimilarity between two sets depending on the type
of their elements. 
Section~\ref{robust} presents a robust privacy-preserving
set intersection protocol that can effectively be used
for implicit authentication.
The privacy, security and complexity of the new protocol
are analyzed in Section~\ref{evaluation}.
Experimental results are reported in Section~\ref{experiment}.
Finally, conclusions and future research directions are
summarized in Section~\ref{conclusion}.
The Appendix gives background on privacy-preserving set intersection, 
recalls the Paillier cryptosystem 
and justifies the correctness of the least obvious steps of
our protocol.

\section{Background}
\label{back}

We first specify the usual setting of implicit authentication
and we then move to privacy-preserving implicit authentication.

\subsection{Implicit authentication}
\label{type}

The usual scenario of implicit authentication is one in which
the user carries a mobile networked device (called just user's device
in what follows) such as a cell phone,
tablet, notebook, etc. The user wishes to authenticate 
to a server in order to use some application. The user may
(or not) use a primary password authentication mechanism.
To strengthen such a primary authentication or even to replace
it, the user resorts to {\em implicit authentication}~\cite{Jako09}.
In this type of authentication, the history of a user's actions
on the user's device is used to construct a profile for the user 
that consists of a set of features. 
In~\cite{Jako09} empirical evidence was given that the features 
collected from the user's device history are effective to 
distinguish users and therefore can be used to
implicitly authenticate them (instead or in addition
to explicit authentication based on the user's providing
a password). 

The types of features collected on the user's actions 
fall into three categories: (i) device data, like GPS 
location data, WiFi/Bluetooth connections and other sensor
data; (ii) carrier data, such as information on cell towers
seen by the device, or Internet access points; 
and (iii) cloud data, such as calendar entries.
It is not
safe to store the accumulated profile of the user in
the user's device,
because an intruder might compromise the device
and alter the stored profile in order  
to impersonate the legitimate user.
Hence, for security, the profile must be stored
by some external entity.
However, the user's profile includes potentially 
sensitive information and storing it outside the user's
device violates privacy.

Implicit authentication systems try to mitigate
the above privacy problem by using a third
party, the {\em carrier} (i.e. the network service provider)
to store the user's profiles. Thus, the typical architecture
consists of the user's device, the carrier and the application servers.
The latter want to authenticate the user and they collaborate
with the carrier and the user's device to do so.
The user's
device engages in a secure two-party computation protocol
with the carrier in order to compare the fresh 
usage features collected by the user's device with 
the user's profile stored at the carrier.
The computation yields a score that is compared
(by the carrier or by the application server) against
a threshold, in order to decide whether the user
is accepted or rejected. In any case, the application
server trusts the score computed by the carrier. 

\subsection{Privacy-preserving implicit authentication}

In the privacy-preserving implicit authentication system
proposed in~\cite{Safa14}, the user's device encrypts
the user's usage profile at set-up time, 
and forwards it to the carrier, who stores it for later
comparison.
There is no security problem because 
during normal operation the user's device does
not store the user's profile
(it just collects the fresh usage features).
There is no privacy problem either, because the carrier
does not see the user's profile in the clear.

The core of proposal~\cite{Safa14} is the algorithm
for computing the dissimilarity score between two inputs: the fresh
sample provided by the user's device and the 
profile stored at the carrier.
All the computation takes place at the carrier and 
both inputs are encrypted: indeed, the carrier
stores the encrypted profile and the user's device
sends the {\em encrypted} fresh sample to the carrier. Note
that the keys to both encryptions are only known to 
the user's device (it is the device who encrypted
everything).

The carrier computes a dissimilarity score at the feature level,
while provably guaranteeing that: i) no information about the 
profile stored at the carrier is revealed to the device
other than the average absolute deviation of the stored
feature values; ii) no information about the fresh feature 
value provided by the device is revealed to the carrier 
other than how it is ordered with respect to the stored
profile feature values.
  
The score computation protocol in~\cite{Safa14} uses
two different encryption schemes: a homomorphic encryption
scheme $HE$ (for example, Paillier's~\cite{Pai}) 
and an order-preserving symmetric encryption scheme
$OPSE$ (for example,~\cite{Bold09}). 
For each feature in the accumulated user's profile, 
two encrypted versions are created, one under $HE$ and 
the other under $OPSE$. Similarly, for each feature
in the fresh sample it collects,
the user's device computes  
two encrypted versions, under $HE$ and
$OPSE$, respectively, and sends them to the carrier.
The following process is repeated for each feature:
\begin{enumerate}
\item Using the $HE$ ciphertexts the carrier performs
some computations (additions and scalar multiplications) relating the  
encrypted fresh sampled feature value
and the set of encrypted feature values in the 
stored encrypted user's profile.
\item~\label{pas2} The output 
of the previous computations is returned to the user's device,
which decrypts it, re-encrypts it under $OPSE$ 
and returns the re-encrypted value to the carrier.
\item Using the order-preserving properties, the carrier
can finally compute a dissimilarity score evaluating 
how different is the fresh
sampled feature from those stored in the encrypted
user's profile. This score can be roughly described
as the number of feature values in the stored encrypted
profile that are less dissimilar from the median 
of the stored values than the fresh sampled value.
\end{enumerate}   

The authors of~\cite{Safa14} point out that,
in case of a malicious user's device ({\em e.g.}
as a result of it being compromised),
one cannot trust the device 
to provide the correct $HE$-encrypted
version of the fresh sampled feature.
Nor can it be assumed
that the device returns correct $OPSE$-encryptions in Step~\ref{pas2}
above. In~\cite{Safa14}, a variant 
of the privacy-preserving implicit authentication protocol
is presented in which the device proves the
correctness of $HE$-encrypted fresh sampled features
and does not need to provide $OPSE$-encrypted values.
This version is secure against malicious
devices, but its complexity is substantially higher.

Other shortcomings of~\cite{Safa14}:
\begin{itemize}
\item It is restricted to numerical features, due to the kind of 
computations that need to be performed on them.
However, among the example features listed in Section~\ref{type},
there are some features that are not numerical,
like the list of cell towers or Internet access points
seen by the user's device.
\item It discloses the following information to the user's device:
i) how the fresh sample is ordered with respect to the stored profile feature
values; ii) the average absolute deviation 
of the stored feature values. 
\end{itemize}

We present a privacy-preserving implicit authentication
protocol based on set intersection that deals with
the above shortcomings. 

\section{Dissimilarity between sets depending on the data type}
\label{dist}

Based on~\cite{Blan14}, we recall here how 
the dissimilarity between two data sets
$X$ and $Y$ can be evaluated using set intersection. 
If we let $X$ be the user's profile and $Y$ be the 
fresh sample collected by the user's device, our 
privacy-preserving implicit authentication setting
presents the additional complication (not present in~\cite{Blan14})
that $X$ is only available in encrypted form
(the carrier stores only the encrypted user's profile).
Anyway, we describe here the 
case of two plaintext sets $X$ and $Y$ and we will deal
with encrypted sets in the following sections.

\subsection{Case A: independent nominal feature values}
\label{s:tA}

Assume $X$ and $Y$ consist of qualitative values, which 
are independent and binary, that is, the relationship
between two values is equality or nothing. 
Take as an example the 
names of the network or phone providers seen by the user's device, 
the operating system run by the device and/or the programs
installed in the device.
In this case, the dissimilarity between
$X$ and $Y$ can be evaluated  
as the multiplicative inverse of the size
of the intersection of $X$ and $Y$, that is
$1/|X \cap Y|$, when the intersection is not empty. If
it is empty, we say that the dissimilarity is $\infty$.

Clearly, the more the coincidences between $X$ and $Y$, 
the more similar is the profile stored 
at the carrier to the fresh sample collected by the device.

\subsection{Case B: correlated categorical feature values}
\label{s:tB}

As in the previous case, we assume the feature values 
are expressed as qualitative features.
However, these may not be independent. For example,
if the feature values are the IDs of cell towers 
or Internet access points seen by the device,
nearby cell towers/access points are more similar
to each other than distant cell towers/access points.

In this case, the dissimilarity between $X$ and $Y$
cannot be computed as the size of their
intersection. 

Assume we have an integer correlation function $l:E\times E\mapsto \Zahl_+$
that measures the similarity between the values in the 
sets of features held by the device and the carrier, where
$E$ is the domain where the sets of features of both
players take values. 
For nominal features, 
semantic similarity measures can be used
for this purpose~\cite{Sanchez12}; for numerical features that take values 
over bounded and discrete domains,
standard arithmetic functions can be used. 
Assume further that both the device and the carrier 
know this function $s$ from the very beginning.

Here the dissimilarity between the set $X$ 
and the set $Y$ can be computed as
\[1/(\textstyle{\sum_{x\in X}\sum_{y\in Y} l(x,y)})\]
when the denominator is nonzero. If it is zero, we say that 
the distance is $\infty$.

\subsection{Case C: numerical feature values}
\label{s:tC}

In this case, we want to compute the dissimilarity between two
sets of numerical values based on set intersection. 
Numerical features in implicit authentication may include
GPS location data, other sensor data, etc.
Assume $U = \{ u_1, \cdots, u_t\}$ and $V = \{v_1, \cdots, v_t\}$.
A way to measure the dissimilarity between $X$ and $Y$ is to compute 
$\sum_{i=1}^{t} |u_i-v_i|$.

\section{Robust privacy-preserving set intersection 
for implicit authentication}
\label{robust}

It will be shown further below that
computing dissimilarities in the above three cases A, B and C
can be reduced to computing the cardinality of set intersections. 
Furthermore, this can be done without the carrier 
revealing $X$ and without the user's device revealing $Y$, as required
in the implicit authentication setting.
The idea is that, if the dissimilarity stays below
a certain threshold, the user is authenticated; otherwise,
authentication is refused.

In Appendix~\ref{setint}, we give some background on 
privacy-preserving set intersection
protocols in the literature.
Unfortunately, all of them assume an honest-but-curious 
situation, but we need a
privacy-preserving set intersection 
protocol that works even if the adversary is a 
malicious one: notice that the user's device may be corrupted, that is,
in control of some adversary. Hence we proceed to specifying 
a set intersection protocol that remains
robust in the malicious scenario and
we  apply it to achieving
privacy-preserving implicit authentication in Case A.
We then extend it to Cases B and C. 
We make use of Paillier's cryptosystem~\cite{Pai}, which is 
recalled in Appendix~\ref{paillier}. 

\subsection{Implicit authentication in Case A}
\label{impA}

\subsubsection{Set-up}
\label{caseA}

Let the plaintext user's profile be $(a_1, \cdots, a_s)$.
In this phase, the user's device transfers the encrypted user's
profile to the carrier. 
To do so, the user's device does:

\begin{enumerate}
  \item Generate the Paillier cryptosystem with public key $pk=(n,g)$ and secret key $sk$.
  \item Compute the polynomial $p(x) = \prod_{i=1}^s (x -a_i)=p_0+p_1x+p_2x^2+\cdots+p_sx^s$.
  \item Compute $Enc(p_0),\cdots Enc(p_s)$ where $Enc(p_i)=g^{p_i}r_i^n \mod n^2$.
  \item Randomly choose $R'\in Z_{n^2}$. Find 
$r_0', \cdots, r_s' \in Z_{n^2}$ such that
 \begin{equation}\label{Vanderm}
  R'=r_0'\cdot {r_1'}^{a_j}\cdot {r_2'}^{a_j^2}\cdots {r_s'}^{a_j^s} \mod n^2, \;\;\; j=1,\cdots,s
 \end{equation}
Note that the system (\ref{Vanderm}) has a trivial solution
$r_0'=R'$ and $r_1'= \cdots = r_s' =1$, but, since it is underdetermined
($s+1$ unknowns and $s$ equations), it has many non-trivial
solutions too (see correctness analysis in Appendix~\ref{correctness}).
  \item Compute $R_i=r_i'/r_i \mod n^2$. Randomly choose integer $d\in Z_{n}$. Send
   $$pk,Enc(p_0),\cdots Enc(p_s);{R_0}^d,\cdots, {R_s}^d \mod n^2$$
   to the carrier. Locally delete all
data computed during the set-up protocol, but keep $(d,R')$ secretly.
\end{enumerate}

\subsubsection{Implicit authentication protocol}

As discussed in Section~\ref{s:tA}, in case of independent
nominal feature values (Case A), dissimilarity is computed as 
$1/|X\cap Y|$. Hence, to perform implicit authentication
the carrier 
just needs to compute the cardinality of the intersection 
between the fresh sample collected by the user's device and the
user's profile stored at the carrier. {\em The challenge is
that the carrier only holds the encrypted user's profile 
and the user's device does no longer hold the plaintext user's profile
either in plaintext or ciphertext.}
 
Let $Y =\{b_1, \cdots, b_t\}\subseteq E$ be the fresh sample collected by the user's device. Then the device and the carrier engage in the following protocol:
\begin{description}
  \item[Step 1] The carrier randomly chooses $\theta$, and sends $pk$, $Enc(p_0)^\theta,$ $\cdots$ $Enc(p_s)^\theta$; ${R_0}^d$, $\cdots$, ${R_s}^d$ to the user's device.

  \item[Step 2] The user's device picks a random integer $r(j)\in \mathbb{Z}_{n^2}$ for every $1\leq j \leq t$. The device computes for $1\leq j \leq t$
\begin{eqnarray}
Enc(r(j)\cdot d \cdot \theta \cdot p(b_j)) &=&Enc(p(b_j))^{d \cdot \theta 
\cdot r(j)}\nonumber\\
                       &=&(Enc(p_0)\cdots Enc(p_s)^{b_j^s})^{d \cdot \theta \cdot r(j)}\nonumber\\
                       &=&g^{r(j) \cdot d \cdot \theta p(b_j)} \gamma_j^{n \cdot d \cdot \theta} \mod n^2\nonumber
\end{eqnarray}
where
$\gamma_j=(r_0\cdot {r_1}^{b_j}\cdot {r_2}^{b_j^2}\cdots {r_s}^{b_j^s})^{r(j)} \mod n^2$.
The user's device then computes
$\Upsilon_j=(R_0\cdot {R_1}^{b_j}\cdot {R_2}^{b_j^2}\cdots {R_s}^{b_j^s})^{dr(j)} \mod n^2$. 
For all $j$, the device randomly orders and sends
\begin{equation}\label{protocoldata}
\{(Enc(r(j)\cdot d \cdot \theta \cdot p(b_j)),\Upsilon_j,{R'}^{r(j)d})\}
\end{equation}
to the carrier.
  \item[Step 3] For $1\leq j \leq t$, the carrier does:
  \begin{itemize}
    \item Compute $Enc(r(j) \cdot d \cdot \theta \cdot p(b_j))\cdot \Upsilon_j^{n\theta}$;
    \item From Expression (\ref{Vanderm}), if $b_j=a_i$
for some $i \in \{1,\cdots, s\}$, then
$p(b_j)=0$ and hence $Enc(r(j)d \cdot \theta \cdot p(b_j))\cdot \Upsilon_j^{n\theta}={R'}^{r(j)dn\theta}$; note that the carrier can recognize
${R'}^{r(j)dn\theta}$ by raising ${R'}^{r(j)d}$ received in Expression (\ref{protocoldata}) to $n \theta$.
Otherwise
(if $b_j\neq a_i$ for all $i \in \{1, \cdots, s\}$) $Enc(r(j)\cdot d \cdot \theta \cdot p(b_j))$ looks random. See correctness analysis in 
Appendix~\ref{correctness}.
  \end{itemize}

\end{description}

If both parties are honest, then the carrier learns $|X\cap Y|$ but obtains no information about the elements in $X$ or $Y$.

\subsection{Implicit authentication in Case B}

Here, the carrier inputs $X$ and the user's device inputs $Y$, 
two sets of features, and they want to know how close $X$ and $Y$ are 
without revealing their own set. 
In the protocol below, only the carrier learns how close $X$ and $Y$ are.

We assume that the domain of $X$ and $Y$ is the same, and
we call it $E$. The closeness or similarity between elements 
is computed by means of a function $s$. In particular, we consider
functions $l:E\times E\to \Zahl_+$. 
Observe that Case A 
is a particular
instance of this Case B in which
$l(x,x)=1$ and $l(x,y)=0$ for $x\neq y$. 

Let $Y$ be the input of the user's device. 
For every $z\in E$, the device computes
$\ell_z=\sum_{y\in Y} l(z,y)$. Observe that $\ell_z$ measures 
the overall similarity of $z$ and $Y$. Let $Y'=\{z\in E\, :\, \ell_z>0\}$. 
It is common to consider functions satisfying $l(z,z)>0$ for every $z\in E$, 
and so in general $Y\subseteq Y'$.

An implicit authentication protocol for such a computation can be obtained from 
the protocol in Case A (Section~\ref{impA}), by replacing Steps 2 and 3 
there with the following ones:

\begin{description}
\item[Step 2'] 
For every $z \in Y'$, the user's device
picks $\ell_z$ random integers $r(1),$ $\cdots,$ $r(\ell_z)$$ \in $$\mathbb{Z}_{n^2}$ 
and for $1 \leq j \leq \ell_z$ does
\begin{itemize}
\item Compute 
\begin{eqnarray}
Enc(r(j)\cdot d \cdot \theta \cdot p(z)) &=&Enc(p(z))^{d \cdot \theta 
\cdot r(j)}\nonumber\\
                       &=&(Enc(p_0)\cdots Enc(p_s)^{z^s})^{d \cdot \theta \cdot r(j)}\nonumber\\
                       &=&g^{r(j) \cdot d \cdot \theta p(z)} \gamma_j^{n \cdot d \cdot \theta} \mod n^2\nonumber
\end{eqnarray}
where
$\gamma_j=(r_0\cdot {r_1}^{z}\cdot {r_2}^{z^2}\cdots {r_s}^{z^s})^{r(j)} \mod n^2$.
\item Compute
$\Upsilon_j=(R_0\cdot {R_1}^{z}\cdot {R_2}^{z^2}\cdots {R_s}^{z^s})^{dr(j)} \mod n^2$.
\item Let $E_j = \{(Enc(r(j)\cdot d \cdot \theta \cdot p(z)),\Upsilon_j,{R'}^{r(j)d})\}$.
\end{itemize} 
Finally, the user's device randomly re-orders the sequence
of all computed $E_j$
for all $z \in Y'$ (a total
of $\sum_{z \in Y'} \ell_z$ elements) 
and sends the randomly re-ordered sequence of $E_j$'s
to the carrier.
\item[Step 3'] For every received $E_j$, the carrier does
  \begin{itemize}
    \item Compute $Enc(r(j)d\theta \cdot p(z))\cdot \Upsilon_j^{n\theta}$;
    \item\label{3b} From Expression (\ref{Vanderm}), if $z \in X$, then
$p(z)=0$ and hence $Enc(r(j)d \cdot \theta \cdot p(z))\cdot \Upsilon_j^{n\theta}={R'}^{r(j)dn\theta}$ (see correctness analysis in Appendix~\ref{correctness}); otherwise
(if $z \not\in X$) $Enc(r(j)d\theta \cdot p(z))$ looks random.
  \end{itemize}
\end{description}

Hence, at the end of the protocol, the total number of $E_j$ 
which yield ${R'}^{r(j)dn\theta}$ is 
\[\sum_{x\in X} \ell_x = \sum_{x\in X}\sum_{y\in Y} l(x,y),\]
that is, the sum of similarities between the elements 
of $X$ and $Y$. This clearly measures how similar 
$X$ and $Y$ are. 
At the end of the protocol, the carrier knows $|Y'|$ and the device knows $|X|$.
Besides that, neither the carrier nor the device
can gain any additional knowledge on the elements of each other's 
set of preferences.

\subsection{Implicit authentication in Case C}
\label{impC}
          
Let the plaintext user's profile be a set $U$
of $t$ numerical features, which we denote by  
$U = \{ u_1, \cdots, u_t\}$. The device's
fresh sample corresponding to those features 
is $V= \{v_1, \cdots, v_t\}$. The carrier wants to 
learn how close $X$ and $Y$ are, that is,
$\sum_{i=1}^{t} |u_i-v_i|$.
                     
Define $X=\{(i,j): u_i > 0 \mbox{ and } 1 \leq j \leq u_i\}$
and $Y= \{(i,j): v_i > 0 \mbox{ and } 1 \leq j \leq v_i\}$.
Now, take the set-up protocol defined in Section~\ref{impA} for Case A and 
run it by using $X$ as plaintext user profile.
Then take the implicit authentication protocol for Case A and 
run it by using $Y$ as the fresh sample input by the device.
In this way, the carrier can compute $|X \cap Y|$.
Observe that
\[ |X\cap Y|=|\{ (i,j) \, :\, u_i,v_i>0 \text{ and } 1\leq j \leq \min\{u_i, v_i\}\}|
=\sum_{1\leq i\leq t}\min\{u_i,v_i\}. \]
In the set-up protocol for Case A, the carrier learns $|X|$
and during the implicit authentication protocol for Case A,
the carrier learns $|Y|$. Hence, 
the carrier can compute
\[ |X|+|Y|-2|X\cap Y|=
\sum_{i=1}^t (\max\{u_i,v_i\}+\min\{u_i,v_i\})-2\sum_{i=1}^t \min\{u_i,v_i\}\]
\[ =\sum_{i=1}^t (\max\{u_i,v_i\}-\min\{u_i,v_i\})
=\sum_{i=1}^t |u_i-v_i| \]
 
\section{Privacy, security and complexity}
\label{evaluation}

Unless otherwise stated, the assessment in this section 
will focus on the protocols of Case A (Section~\ref{impA}),
the protocols of Cases B and C being extensions of Case A. 

\subsection{Privacy and security}

We define privacy in the following two senses:
\begin{itemize}
\item After the set-up is concluded, the user's device does not 
keep any information about the user's profile sent
to the carrier. Hence, compromise
of the user's device does not result in compromise of the user's profile.
\item The carrier learns nothing about the plaintext user's profile,
except its size. This
allows the user to preserve the privacy of her profile towards the 
carrier.
\end{itemize}

\begin{lemma}
\label{lem1}
After set-up, the user's device does not keep any information on
the user's profile sent to the carrier.
\end{lemma}

{\bf Proof.} The user's device only keeps $(d,R')$ at the end
of the set-up protocol. Both $d$ and $R'$ are random and hence
unrelated to the user's profile. $\Box$.

\begin{lemma}
The carrier or any eavesdropper 
learn nothing about the plaintext user's profile,
except its size.
\end{lemma}

{\bf Proof.} After set-up, the carrier receives
 $pk,Enc(p_0),\cdots Enc(p_s)$;${R_0}^d,$ $\cdots,$ ${R_s}^d \mod n^2.$ 
Since $d$ is random and unknown to the carrier, 
${R_0}^d,\cdots, {R_s}^d \mod n^2$ look random to the carrier and 
will give him no more information about the plaintext user's profile
than the Paillier ciphertexts $Enc(p_0),\cdots Enc(p_s)$. That is,
the carrier learns nothing about the user's plaintext
profile $X=\{a_1,$ $\cdots,$ $a_s\}$ except its size $s$.
The same holds true for an eavesdropper listening to the 
communication between the user's device and the carrier during set-up.

At Step 2 of implicit authentication, the carrier only gets the fresh
sample $Y$ encrypted under Paillier and randomly re-ordered. 
Hence, the carrier learns no information on $Y$, except its size $t$.
At Step 3, the carrier learns $|X \cap Y|$, but not knowing 
$Y$, the size $|X \cap Y|$ of the intersection leaks to him no information
on $X$. $\Box$

If we define security of implicit authentication as the inability 
of a dishonest user's device to disrupt the authentication outcome,
we can state the following result.

\begin{lemma}
A dishonest user's device has no better strategy to 
alter the outcome of implicit 
authentication than trying to randomly guess
the user's profile. 
\end{lemma} 

{\bf Proof.} At the end of the set-up protocol, the 
(still uncompromised) user's keeps no information 
about the user's profile (Lemma~\ref{lem1}).
Hence, if the user's device is later compromised and/or behaves
dishonestly, it still has no clue on the real user's profile against
which its fresh sample is going to be authenticated. Hence, either
the user's device provides an honest fresh sample and implicit
authentication will be correctly performed, or the user's device
provides a random fresh sample with the hope
that it matches the user's profile.
$\Box$

\subsection{Complexity}

\subsubsection{Case A}

During 
the set-up protocol, the user's device needs to compute:
\begin{itemize}
\item $s+1$ Paillier encryptions for the polynomial coefficients;
\item values $r'_0, \cdots, r'_s$; as explained in 
Appendix~\ref{correctness}, this can be done by randomly choosing $r'_0$, 
then solving an $s \times s$ generalized Vandermonde system
(doable in $O(s^2)$ time
using~\cite{Demm05}) and finally computing $s$
modular powers to find the $r'_1, \cdots, r'_s$;
\item $s+1$ modular powers (raising the $R_i$ values to $d$).
\end{itemize}

During the implicit authentication protocol, the user's device 
needs to compute (Step 2):
\begin{itemize}
\item $t$ Paillier encryptions;
\item $ts$ modular powers (to compute the $\Upsilon_j$ values);
\item $t$ modular powers (to raise $R'$ to $r(j)d$).
\end{itemize}

Also during the implicit authentication protocol, the carrier
needs to compute:
\begin{itemize}
\item At Step 1, $s+1$ modular powers (to raise
the encrypted polynomial coefficients to $\theta$);
\item At Step 3, $t$ Paillier encryptions;
\item At Step 3, $t$ modular powers (to raise the 
$\Upsilon_j$ values to $n\theta$).
\end{itemize} 

\subsubsection{Case B}

The set-up protocol does not change w.r.t. Case A. 
In the implicit authentication protocol, the highest
complexity occurs when $Y' = E$ and the similarity
function $l$ always takes the maximum value in its range, say $L$. 
In this case, 
$$\sum_{z \in Y'} \ell_z = \sum_{z \in Y'} \sum_{y \in Y} l(z,y)
= |E| s L.$$
Hence, in the {\em worst case} the user's
device needs to compute (Step 2'):
\begin{itemize}
\item $|E|sL$ Paillier encryptions;
\item $|E|sL$ modular powers (to 
compute the $\Upsilon_j$ values); 
\item $|E|sL$ modular powers
(to raise  $R'$ to $r(j)d$).
\end{itemize}

Also during the implicit authentication protocol, the carrier
needs to compute:
\begin{itemize}
\item At Step 1, $s+1$ modular powers (to raise
the encrypted polynomial coefficients to $\theta$);
\item At Step 3', $|E|sL$ Paillier encryptions;
\item At Step 3', $|E|sL$ modular powers (to raise the 
$\Upsilon_j$ values to $n\theta$).
\end{itemize} 

Note that the above complexity can be reduced by 
reducing the range of the similarity function $l(\cdot,\cdot)$.

\subsubsection{Case C}

Case C is analogous to Case A but the sets $X$ and $Y$ whose intersection
is computed no longer have $s$ and $t$ elements, respectively.
According to Section~\ref{impC}, the 
maximum value for $|X|$ occurs when all $u_i$ take the maximum
value of their range, say, $M$, in which case 
$X$ contains $tM$ pairs $(i,j)$. By a similar argument,
$Y$ also contains at most $tM$ pairs.

Hence, the {\em worst-case} complexity for Case C is obtained
by performing the corresponding changes in the assessment
of Case A. Specifically, during
the set-up protocol, the user's device needs to compute:
\begin{itemize}
\item $tM+1$ Paillier encryptions for the polynomial coefficients;
\item Solve a Vandermonde system $tM\times tM$
(doable in $O((tM)^2)$ time) and then 
compute $tM$ modular powers to find the $r'_i$ values;
\item Compute $tM+1$ modular powers (raising the $R_i$ values to $d$).
\end{itemize}

During the implicit authentication protocol, the user's device
needs to compute (Step 2):
\begin{itemize}
\item $tM$ Paillier encryptions;
\item $t^2M^2$ modular powers (to compute the $\Upsilon_j$ values);
\item $tM$ modular powers (to raise $R'$ to $r(j)d$).
\end{itemize}

Also during the implicit authentication protocol, the carrier
needs to compute:
\begin{itemize}
\item At Step 1, $tM+1$ modular powers (to raise
the encrypted polynomial coefficients to $\theta$);
\item At Step 3, $tM$ Paillier encryptions;
\item At Step 3, $tM$ modular powers (to raise the
$\Upsilon_j$ values to $n\theta$).
\end{itemize}

Note that the above complexities can be reduced by reducing
the range of the numerical values in sets $U$ and $V$.

\section{Experimental results}
\label{experiment}

As stated in the previous section, the complexity of our 
implicit authentication
protocol ultimately depends on the sizes of the input sets. 
In Case A, the sizes of
the sets are directly given by the user inputs; in Case B, 
these sizes are the product
of the size of the input sets times 
the range of the similarity function $\ell$; and in
Case C, the sizes are given by the size of 
the original sets times the range of their
values. We ran an experiment to test the execution times of our protocol, based 
on Case A, to which the other two cases can be reduced. 

The experiment was implemented in Sage-6.4.1 and 
run on a Debian7.7 machine with 
a 64-bit architecture, an Intel i7 processor and 8GB of physical memory. 
We instantiated a Paillier cryptosystem 
with a 1024-bit long $n$, and the 
features of preference sets were taken from the integers in the range $[1 \dots 2^{128}]$.
The input sets ranged from size $1$ to $50$, and we took 
feature sets of the same 
size to execute the set-up and the authentication protocols.

Step 4 of the set-up protocol (Section~\ref{caseA}), 
in which a system of equations is solved 
for $r'_i$ for $1 \le i \le s$, is the most expensive part of the 
set-up protocol. 
As a worst-case setting, we used straightforward Gaussian elimination 
which takes time $O(s^3)$, although, as mentioned above,
specific methods like~\cite{Demm05} exist for generalized Vandermonde matrices 
that can run in $O(s^2)$ 
(such specific methods could be leveraged in case 
of smartphones with low computational power).
On the other hand, Step 2 of the authentication protocol
(Section~\ref{caseA}), 
computed by the user's device, is easily parallelizable for each feature 
in the sample set. Since parallelization can be exploited
by most of the current 
smartphones in the market, we also exploited it in our experiment.
The results are shown in Table~\ref{exeTimes} (times are in seconds).


\begin{table}
\centering
\caption{Execution times (in seconds) for different input set sizes}
\label{exeTimes}
\begin{tabular}{|c|c|c|c|c|c|c|c|c|c|c|c|}
\hline 
 & 1 & 5 & 10 & 15 & 20 & 25 & 30 & 35 & 40 & 45 & 50 \\
\hline
Set-up & 0.89 & 0.79 & 1.1 & 1.83 & 4.67 & 11.45 & 24.65 & 47.6 & 84.99 & 144.81 & 228.6 \\
\hline
Authentication & 0.08 & 0.47 & 1.05 & 2.0 & 3.37 & 5.4 & 8.27 & 12.13 & 17.3 & 23.39 & 31.2 \\
\hline
\end{tabular}
\end{table}

Note that the set-up protocol is run only once 
(actually, maybe once in a while), so it is not time-critical.
However, the authentication protocol is to
be run at every authentication attempt by the user.
For example, if a user implicitly authenticates herself 
using the pattern of her 20 most visited websites, authentication
with our proposal would take 3.37 seconds, which is perfectly 
acceptable in practice.

\section{Conclusions and future research}
\label{conclusion}

To the best of our knowledge, we have presented the 
second privacy-preserving implicit authentication system
in the literature (the first one was~\cite{Safa14}).
The advantages of our proposal with respect to~\cite{Safa14} are:
\begin{itemize}
\item The carrier only needs to store the user's profile 
encrypted under {\em one} cryptosystem, namely Paillier's.
\item Dishonest behavior or compromise at the user's device
after the initial set-up stage neither compromises the privacy
of the user's profile nor affects the security of authentication.
\item Our proposal is not restricted
to numerical features, but can deal also with all sorts
of categorical features.
\item In case of numerical or categorical ordinal features,
our proposal does not disclose how the fresh sample 
is ordered with respect to the feature
values in the stored user's profile.
\end{itemize}

For binary or independent nominal features, the complexity
of our proposal is quite low (quadratic in the number of 
values in the user's profile). 
For correlated categorical feature values, the complexity
is higher, but it can be reduced by decreasing the range 
of the similarity function used. Finally, in the case 
of numerical values, the complexity is also higher
than in the binary/independent nominal case, but it can
be reduced by decreasing the range of the numerical feature values.

Future research will include devising ways to further decrease
the computational complexity in all cases.

\section*{Acknowledgments}

The following funding sources are gratefully acknowledged:
Government of Catalonia (ICREA Acad\`emia Prize to the
first author and grant 2014 SGR 537),
Spanish Government (project TIN2011-27076-C03-01 ``CO-PRIVACY''),
European Commission (projects
FP7 ``DwB'', FP7 ``Inter-Trust'' and H2020 ``CLARUS''),
Templeton World Charity
Foundation (grant TWCF0095/AB60 ``CO-UTILITY''),
Google (Faculty Research Award to the first author) and
Government of China (Natural Science Foundation of China
under projects 61370190 and 61173154).
The first author is with the UNESCO Chair in Data Privacy.
The views in this paper are the authors' own and
do not necessarily reflect
the views of UNESCO, the Templeton World Charity
Foundation or Google.

\bibliographystyle{plain}

\appendix

\section{Background on privacy-preserving set intersection}
\label{setint}

Secure multiparty computation (MPC) allows a set of parties to compute
functions of their inputs in a secure way without requiring a trusted 
third party. During the execution of the protocol, the parties do 
not learn anything about each other's input except what is implied by the 
output itself. 
There are two main adversarial models: honest-but-curious adversaries 
and malicious adversaries. In the former model, the parties follow 
the protocol instructions but they try to obtain information 
about the inputs of other parties from the messages they receive. 
In the latter model, 
the adversary may deviate from the protocol in an arbitrary way. 

We will restrict here to a two-party setting in which
the input of each party is a set, and the desired output is the
cardinality of the intersection of both sets. 
The intersection of two sets can be obtained by using generic constructions based on Yao's garbled circuit~\cite{Yao86}. This technique 
allows computing any arithmetic function, but for most of the functions it is inefficient. Many of the recent works on two-party computation are focused on improving the efficiency of these protocols for particular families of functions.

Freedman, Nissim, and Pinkas~\cite{FNP} presented a more efficient meth\-od to compute the set intersection, a {\em private matching scheme}, that is secure in the honest-but-curious model. A private matching scheme is a protocol between a client $\alice$ and a server $\bob$ in which $\alice$'s input is a set $X$ of size $i_{\alice}$, $\bob$'s input is a set $Y$ of size $i_{\bob}$, and at the end of the protocol $\alice$ learns $X\cap Y$. The
scheme uses polynomial-based techniques and homomorphic encryption schemes. 
Several variations of the private matching scheme were also presented in~\cite{FNP}: an extension to the malicious adversary model, an extension of the multi-party case, and schemes to compute the cardinality of the set intersection and other functions. Constructing efficient schemes for set operations is an important topic in MPC and has been studied in many other 
contributions. Several works such as~\cite{BlAg,DGT,HoWe,KiSo,Vaidya} present new protocols to compute the set intersection cardinality.

\section{Paillier's cryptosystem}
\label{paillier}

In this cryptosystem, the public 
key consists of an integer $n$ (product of two 
RSA primes), and an integer $g$ of order $n$ modulo $n^2$, for 
example, $g=1+n$. The secret key is $\phi(n)$, where $\phi(\cdot)$
is Euler's totient function.

Encryption of a plaintext integer $m$, with $m<n$ 
involves selecting a random integer $r < n$ and computing
the ciphertext $c$ as 
\[ c = Enc(m) = g^m \cdot r^n \bmod n^2 = (1+mn)r^n \bmod n^2. \] 
Decryption consists of first computing 
$c_1 = c^{\phi(n)} \bmod n^2 = 1 + m \phi(n) n \bmod n^2$ 
and then 
$m = (c_1 -1) \phi(n)^{-1} \bmod n^2$. 

The homomorphic properties of Paillier's cryptosystem
are as follows:
\begin{itemize}
\item {\em Homomorphic addition of plaintexts}. The product of two ciphertexts
decrypts as the sum of their corresponding plaintexts:
\[ D(E(m_1,r_1) \cdot E(m_2,r_2) \bmod n^2) = m_1 + m_2 \bmod n.\]
Also, the product of a ciphertext times $g$ raised to a plaintext
decrypts as the sum of the corresponding plaintexts:
\[ D(E(m_1,r_1) \cdot g^{m_2} \bmod n^2) = m_1+m_2 \bmod n. \]
\item {\em Homomorphic multiplication of plaintexts}. An encrypted
plaintext raised to the power of another plaintext will decrypt to 
the product of the two plaintexts:
\[ D(E(m_1, r_1)^{m_2} \bmod n^2) =  D(E(m_1, r_1)^{m_2} \bmod n^2) = m_1 m_2 \bmod n. \]
More generally, given a constant $k$,
$D(E(m_1, r_1)^k \bmod n^2) = k m_1 \bmod n$. 
\end{itemize}

\section{Correctness}
\label{correctness}

In general, the correctness of our protocol follows 
from direct algebraic verification
using the properties of Paillier's cryptosystem.
We go next through the least obvious steps.

\subsection{Set-up protocol}

In the set-up protocol, $r'_0,\cdots,r'_s$ are found as a solution
of the following system

\[\left[\begin{array}{c} R'\\ \vdots \\R' \end{array}\right]
= \left[\begin{array}{c} r'_0 \cdot {r'}_1^{a_1} \cdot {r'}_2^{a_1^2} \cdots {r'}_s^{a_1^s} \mod n^2\\
\vdots \\
r'_0 \cdot {r'}_1^{a_s} \cdot {r'}_2^{a_s^2} \cdots {r'}_s^{a_s^s} \mod n^2
\end{array} \right].\]

The above system has $s+1$ unknowns and $s$ equations. Therefore
it has one degree of freedom. To avoid the 
trivial solution
$r_0'=R'$ and $r_1'= \cdots = r_s' =1$, we choose a random $r_0'$.
Then we divide the system by $r_0'$ and we take logarithms
to get 
\[ \left[\begin{array}{c} \log (R'/r'_0)\\ \log (R'/r'_0)\\ \vdots \\\log (R'/r'_0) \end{array}\right] \mod n 
= \left[\begin{array}{cccc}   
a_1 & a_1^2 & \cdots & a_1^s\\
\vdots & \vdots & \vdots\\
a_s & a_s^2 & \cdots & a_s^s
\end{array} \right] \cdot 
\left[\begin{array}{c} \log r'_1 \\ \log r'_2 \\ \vdots \\\log r'_s \end{array}\right] \mod n. \]
The matrix on the right-hand side of the above system
is an $s \times s$ generalized Vandermonde matrix (not quite a Vandermonde
matrix). Hence, using the techniques in~\cite{Demm05} 
it can be solved in $O(s^2)$ time
for $\log r'_1$, $\cdots$, $\log r'_s$.
Then $s$ powers modulo $n^2$ need to be computed
to turn $\log r'_i$ into $r'_i$ for $i=0,\cdots,s$.

\subsection{Implicit authentication protocol}

We specify in more detail 
the following derivation in Step 2 of the implicit authentication protocol of 
Section~\ref{impA}:
\begin{eqnarray}
Enc(r(j)\cdot d \cdot \theta \cdot p(b_j)) &=&Enc(p(b_j))^{d \cdot \theta 
\cdot r(j)} \mod n^2\nonumber\\
                       &=&(Enc(p_0)\cdots Enc(p_s)^{b_j^s})^{d \cdot \theta \cdot r(j)} \mod n^2\nonumber\\
		      &=& (g^{p_0} r_0^n \cdots (g^{p_s} r_s^n)^{b_j^s})^{d \cdot \theta \cdot r(j)}\mod n^2\nonumber\\
		      &=& (g^{p(b_j)})^{d \cdot \theta \cdot r(j)} (r_0 \cdot r_1^{b_j} \cdots r_s^{b_j^s})^{r(j) \cdot n \cdot d \cdot \theta} \mod n^2\nonumber \\  	
                       &=&g^{r(j) \cdot d \cdot \theta p(b_j)} \gamma_j^{n \cdot d \cdot \theta} \mod n^2\nonumber.
\end{eqnarray}

Regarding Step 3 of the implicit authentication protocol, we detail
the case $b_j=a_i$ for some $i \in \{1, \cdots, s\}$. 
In this case, $p(b_j)=0$ and hence
\[ Enc(r(j)d \theta \cdot p(b_j))\cdot \Upsilon_j^{n\theta} \mod n^2  = 
Enc(0)^{r(j)d \theta} \cdot \Upsilon_j^{n\theta} \mod n^2 \]
\[ = (r_0 \cdot r_1^{b_j} \cdots r_s^{b^s_j})^{n r(j) d \theta} \cdot \Upsilon_j^{n\theta} \mod n^2 \] 
\[ = (r_0 \cdot r_1^{b_j} \cdots r_s^{b^s_j})^{n r(j) d \theta} \cdot
(R_0 \cdot R_1^{b_j} \cdots R_s^{b^s_j})^{dr(j)n\theta} \mod n^2 \]
\begin{equation}
\label{deriv}
 = ({r'}_0 \cdot {r'}_1^{a_i} \cdots {r'}_s^{a^s_i})^{r(j) d n \theta} \mod n^2 = {R'}^{r(j)d n \theta} \mod n^2. 
\end{equation} 

If in Step 3, if we have $b_j \neq a_i$ for all $i \in \{1,\cdots,s\}$, then
Derivation (\ref{deriv}) does not hold and 
a random number is obtained instead. On the one side,
the powers of $g$ does not disappear from $Enc(r(j)d \theta \cdot p(b_j))$.
On the other side, the 
exponents $b_j, \cdots, b_j^s$ cannot be changed by
$a_i, \cdots, a_i^s$ as done in the last step of Derivation (\ref{deriv}).
Hence, a random number different from $R'^{r(j)dn\theta}$ is obtained.

\end{document}